\documentclass[preprint]{aastex}

\slugcomment{Draft for the Astrophysical Journal, \today}

\begin{document}

\title {The Formation Rate of Blue Stragglers in 47 Tucanae}
\author{Alison Sills}
\affil{Department of Astronomy, The Ohio State University, 140 W. 18th
Ave., Columbus, OH, 43210}
\email{asills@astronomy.ohio-state.edu}
\author{Charles D. Bailyn}
\affil{Department of Astronomy, Yale University, P. O. Box 208101, New
Haven, CT, 06520-8101}
\email{bailyn@astro.yale.edu}
\author{Peter D. Edmonds} 
\affil{Harvard College Observatory, 60 Garden Street, Cambridge, MA 02138}
\email{pedmonds@cfa.harvard.edu}
\author{Ronald L. Gilliland}
\affil{Space Telescope Science Institute, 3700 San Martin Drive,
Baltimore, MD, 21218}
\email{gillil@stsci.edu}

\begin{abstract}
We investigate the effects of changes in the blue straggler formation
rate in globular clusters on the blue straggler distribution in the
color-magnitude diagram. We find that the blue straggler distribution
is highly sensitive to the past formation rate. Comparing our models
to new UBV observations of a region close to the core of 47 Tucanae
suggests that this cluster may have stopped forming blue straggler
formation several Gyr ago. This cessation of formation can be
associated with an epoch of primordial binary burning which has been
invoked in other clusters to infer the imminence of core collapse.
\end{abstract}

\keywords{blue stragglers -- globular clusters -- 47 Tucanae}

\section{INTRODUCTION}

Blue stragglers in globular clusters are thought to be
created by stellar mergers.  Such mergers can occur in two ways:
through the spiraling in and merger of two components of a binary
system, or through the direct collision of two stars.  The former
mechanism is not strongly dependent on cluster density, but the latter
occurs more often as the stellar collision rate increases.  In a
cluster of single stars, the collision rate is a function of cluster
density and velocity dispersion \citep{vh87}, but a significant binary
population can increase the collision rate well beyond that of a
cluster of single stars.  The enhanced collisions are caused by
resonant encounters with binary stars, which create many more
opportunities for the stars involved to collide, thus greatly
increasing the collisional cross-section \citep{L89}.  Thus the
formation rate of collisional blue stragglers depends on the current
and past cluster density profile, velocity dispersion, and binary
population.  By studying the number of blue stragglers and their
distribution in the color magnitude diagram, we can therefore hope to
probe the dynamical history and stellar populations of the cluster.

HST observations of the cores of globular clusters, combined with
models of blue straggler formation, have been used to infer global
properties of clusters \citep{BP95,SB99,OP}.  These studies suggest
that blue stragglers in the cores of dense clusters are indeed
collisional in origin, and place limits on the binary fraction, mass
function, central density, and velocity dispersion of the clusters.
Recently, \cite{fer99} found a remarkably high blue straggler
concentration in M80, which was difficult to explain given the
relatively low inferred collision rate.  A similar situation pertains
in M3, although it is less pronounced \citep{fer97}.  Ferraro et
al. therefore suggested that M80, and possibly M3, may be in an
unusual dynamical state, in which the density has recently become
large enough to create a large number of encounters involving
primordial binaries, engendering anomalously large collision rates.
Such a state may also be required to explain the anomalous, and
probably short lived, remnants in the core of NGC~6397
\citep{C98,E99}.  The high central density of binaries in NGC 6752
\citep{RB97} may also imply that the cluster is in an unusual
dynamical phase. Once the initial population of primordial binaries
has been "burned", the collision rate would then be expected to
decrease, even as the cluster density continues to rise. Since the
primordial binary burning phase is presumably short, it is somewhat
disturbing that such a phase has to be invoked at the current time in
several different clusters.

In our previous exercises in blue straggler population synthesis, we
have assumed approximately constant collision rates.  Here, we explore
the effects of significant changes in the blue straggler formation
rate on the observed distribution of blue stragglers in the
color-magnitude diagram.  We find that the currently observed blue
straggler populations should vary significantly depending on the past
formation rate.  We apply our results to a new data set from 47~Tuc.
The results suggest that this cluster may well have undergone a burst
of blue straggler formation which ended several Gyr in the past.  In
section 2 we present the theoretical models of blue straggler
distributions. We discuss the observations of 47~Tuc in section 3, and
compare the theory with these observations in section 4. We summarize
our findings in section 5.

\section{MODELS OF BLUE STRAGGLER DISTRIBUTIONS}

We calculate blue straggler distributions in the color-magnitude
diagram (hereafter CMD) of 47 Tuc following the method described in
detail in \cite{SB99}.  We assume that the blue stragglers are all
formed through stellar collisions between single stars during an
encounter between a single star and a binary system.  The trajectories
of the stars during the collision are modeled using the STARLAB
software package \citep{MH96}. The masses of the stars involved are
chosen randomly from a mass function for the current cluster and a
different mass function which governs the mass distribution within the
binary system.  A binary fraction, and a distribution of semi-major
axes must also be assumed.  The output of these simulations is the
probability that a collision between stars of specific masses will
occur. We have chosen standard values for the mass functions and
binary distribution. The current mass function has an index $x=-2$,
and the mass distribution within the binary systems are drawn from a
Salpeter mass function ($x=1.35$). We chose a binary fraction of 20\%
and a binary period distribution which is flat in log P. The effect of
changing these values is explored in \cite{SB99}.  The collision
products are modeled by entropy ordering of gas from colliding stars
\citep{SL97} and evolved from these initial conditions using the Yale
stellar evolution code YREC \citep{G92}. The models reported here used
a metallicity appropriate for 47~Tuc, but the general features we
report are similar for any metallicity. By weighting the resulting
evolutionary tracks by the probability that the specific collision
will occur, we obtain a predicted distribution of blue stragglers in
the CMD.

In order to explore the effects of non-constant blue straggler
formation rates, we examined a series of truncated rates.  In these
models we assumed that the blue straggler formation rate was constant
for some portion of the cluster lifetime, and zero otherwise.  This
assumption is obviously unphysical --- the relevant encounter rates
would presumably change smoothly on timescales comparable to the
relaxation time.  However these models do demonstrate how the
distribution of blue stragglers in the CMD depend on when the blue
stragglers were created, and thus provide a basis for understanding
more complicated and realistic formation rates.

In Figure 1 we show blue straggler distributions in the CMD for
formation rates which were initially zero, and then abruptly
``switched on'' at some point in the cluster's past, and continued at
a constant rate until the present day.  The first panel is the
limiting case of constant formation rate throughout the cluster's
lifetime.  There are dramatic changes as the onset of blue straggler
formation moves closer to the present.  In particular, the redder blue
stragglers disappear, starting from the faint end, until in Figure 1E
the lower part of the blue straggler distribution closely approximates
the zero age main sequence (ZAMS).

This behavior is straightforward to interpret.  Lower mass blue
stragglers start out essentially as ZAMS stars --- they are generally
formed from low mass precursors which have not processed significant
amounts of nuclear fuel, so they have no chemical anomalies.  Since
their main sequence lifetimes are $>>1$ Gyr, the bottom of the
sequence of recently formed blue stragglers closely approximates the
ZAMS.  In contrast, the more massive blue stragglers evolve much
faster, and they are also formed far from the ZAMS in the first place,
since their precursors have already undergone considerable nuclear
processing.  Thus a burst of recent blue straggler formation will
create a blue straggler distribution like that in Figure 1E, with a
narrow sequence at the low L end, and a relatively large number of
stars with a range of temperatures at the bright end.

Figure 2 shows a sequence of blue straggler distributions in which
blue straggler formation began at the start of the cluster lifetime,
but terminated at some point in the past.  The limiting case when the
termination point is the present is the same as Figure 1A and has thus
been omitted.  Figure 2 shows progressively older blue straggler
sequences.  Once again, the dramatic changes in distribution are easy
to understand.  The more massive and luminous blue stragglers evolve
first, and move away from the ZAMS, and then out of the blue straggler
region altogether when they become giants.  A population of blue
stragglers like that shown in Figure 2d, in which all of the blue
stragglers have ages $\ge 8$Gyrs, will therefore contain only
relatively faint blue stragglers and will be skewed toward the red
away from the ZAMS. The dramatic difference between Figure 1E and
Figure 2D, which were created using identical assumptions about binary
fraction, mass function, and other dynamical parameters, illustrates
the importance of including changes in formation rate in studies of
blue straggler distributions. It is difficult to produce such drastic
changes in the shape of the blue straggler distribution by varying the
mass functions and binary fraction, although these parameters do have
a strong influence on the total number of blue stragglers
\citep{SB99}.

Fig 3 shows distributions of blue stragglers in which the formation
rate turned on at some point after the cluster was born, and then
turned off again prior to the present.  As might be expected, these
distributions show characteristics similar to those in both Figs 1 and
2, since both effects described above apply in these cases. We have
also used the binary destruction rate from Figure 3 of \cite{hmr92} as
an approximation for blue straggler creation, since both effects
result from the same close stellar encounters.  The resulting
distribution is dominated by old, low luminosity blue stragglers, but
also contains a small, but potentially observable population of
younger blue stragglers (Figure 4).

\section{OBSERVATIONS OF BLUE STRAGGLERS IN 47 TUCANAE}

Observations of 47 Tucanae were obtained between July 1996 and January
1997.  Data were taken nearly every night for 6 weeks, with some
additional coverage over six months with the CTIO 0.9 m telescope and
2K CCD. Repeated UBVI images were obtained for a 13' $\times$ 13'
field centered on RA,DEC (2000) = 00:22:06.75 -72:04:22.1, with the
closest edge 138.5'' west of cluster center.  The primary purpose was
to study variability on the giant branch, and the time series results
will be presented elsewhere. In this paper, we present color-magnitude
diagrams created from summed data. The exposure times were chosen to
avoid saturation of giant branch stars, and are therefore deepest in
bluer bandpasses, which makes this data set ideal for studying hot
stars, such as blue stragglers. The summed images were analyzed with
DAOPHOT and calibrated with Landolt standards. The calibration agrees
with that of \cite{hess87} to within 1\% for B and V. The stars
presented and analyzed in this paper are only those which contribute
$> 50\%$ of the light within one PSF radius of their centers. This
criterion results in the loss of many crowded stars, especially at or
below the main sequence turnoff.  However the principle sequences
derived are quite clean, and the completeness above the turnoff is
high, though not 100\%.  Figure 5 presents the resulting CMD.

In order to study the blue stragglers, we must have a consistent way
of selecting them from the color-magnitude diagram. It is necessary to
make the selection in two colors, since some stars which are present
in the blue straggler region in one color may show up as photometric
anomalies in other colors. These stars could be foreground or
background objects, photometric errors, or other kinds of strange
stars which are not blue stragglers. The initial selection was done in
the U, U-B diagram (see Figure 5). An additional selection made in V,
B-V diagram (see Figure 6), and then stars far from the principle
sequence in the color-color diagram were rejected (see Figure
7). Although some of the stars excluded may still be blue stragglers,
we have adopted these criteria so that we can have a clean comparison
of the data to our theoretical distributions. The star at V$\sim$14.7,
B-V$\sim$0.2 and U-B$\sim$-0.05 is known to be a variable star
(Edmonds 1999, private communication) and is likely an SX Phoenicis
star. However, it does not satisfy our selection criteria, and
therefore has been rejected from our sample. Using these criteria, we
find 61 blue stragglers (compared to the 20 found by \cite{dem93}). It
should be noted that some of the blue stragglers within 0.75
magnitudes above the cluster turnoff could result from the
superposition of main sequence stars, either by chance or from being a
physical binary. The blue straggler frequency relative to horizontal
branch stars (as defined by \cite{fer99}) is 0.37.  We matched our
theoretical distributions of BSs to the observations by forming our
distributions from those parts of the evolutionary paths which
satisfied the above observational selection criteria. We chose to use
this data set alone, rather than combining it with the earlier HST
data on blue stragglers from the core of the cluster. In order to have
a convincing comparison of theory to data, we need a consistent way of
selecting the blue stragglers, which can be done best with a large
homogeneous data set. In order to understand the properties of 47 Tuc
as a whole, eventually data from all sources and all regions of the
cluster will have to be considered. The implications of our choice
will be discussed in the following section.

\section{COMPARISON OF THEORY WITH OBSERVATIONS}

The theoretical blue straggler distribution with a constant blue
straggler formation rate is shown in Figure 8, along with the 61
selected blue stragglers. This distribution does not match the
observations in three important ways. Firstly, the models predict a
large peak of low luminosity blue stragglers which is not
observed. This is likely a selection effect, since fainter stars are
less likely to pass the 50\% contamination test noted above.
Secondly, there are too many observed blue stragglers at the red end
of the distribution. These so-called ``yellow stragglers'' have been
noted as anomalies before \citep{stet94} and may be due to the
composite colors of binary stars, or chance superpositions which are
fit by only one star in the reduction procedure. Both of these
suggestions can be well studied with simulations of the completeness
and crowding effects, and will be discussed in a future paper.
Thirdly, the theory predicts too many bright blue stragglers. Since
the first two problems cannot be addressed in the context of our
theoretical models, we focus here on the third point, and explore what
is required to produce a theoretical blue straggler distribution which
terminates at the same magnitudes as the observed blue stragglers.

As discussed above, the bright blue stragglers have high masses, and
do not live very long. Therefore, in order to have a population of
blue stragglers which lacks bright stars, the blue stragglers must
have stopped forming some time ago.  In the context of the models
described above, we find that a blue straggler formation rate which
terminated 3 Gyr ago reproduces the upper part of the observed blue
straggler distribution quite well (Figure 9).  However, we caution
that the precise date of the termination of blue straggler formation
should not be taken too seriously.  First, the formation rates used
here are not realistic.  A full dynamical model of the evolution of
the cluster would be required to produce accurate time-dependent
rates.  Second, our results are influenced by our choice of binary
parameters and mass functions, although the same qualitative effects
will apply regardless of the choice of these parameters.  Third, the
observed sample is biased in two important ways.  Incompleteness due
to crowding will affect the distribution, particularly at the faint
end. However this should not affect the lack of bright blue
stragglers, which is the observed feature we are trying to reproduce.
More importantly, we do not have complete spatial coverage of the
cluster.  HST results suggest that the blue straggler distribution
extends to brighter limits in the cluster core \citep{Getal98}. It is
possible that the blue straggler distribution is different in the core
because the contribution of blue stragglers created by binary mergers,
rather than stellar collisions, is larger in the outer regions. If so,
the lack of bright blue stragglers in this region may be more closely
related to the characteristics of the binary population in this region
than the stellar collision rate. Detailed models of binary merger
evolutionary tracks, combined with predictions of the binary
population's merger rate, will be necessary to untangle this
degeneracy.  The lack of bright blue stragglers outside the core could
be explained by mass segregation, either because the more massive blue
stragglers sink to the cluster center (although this effect should not
be dominant since the mass difference between the bright and faint
blue stragglers would be relatively small), or by driving the few
remaining binaries toward the center of the cluster.  However, we do
not believe that mass segregation alone could account for the sharp
cutoff in blue straggler luminosities that we observe in the absence
of a significant change in the blue straggler formation rate, since it
is hard to believe the upper part of the blue straggler distribution
could be lost from our observations given that there are large numbers
of observed blue stragglers in the $1.1-1.4 M_{\sun}$ range.

Thus the data appear to suggest that 47~Tuc has passed through a stage
similar to the current state of M80 at some point in the past.  The
large extent of the blue straggler sequence in M80 observed by
\cite{fer99} tends to support this interpretation, since a cluster
whose blue straggler formation rate is unusually high at the present
time should tend to appear like that in Figure 1D and 1E. Extending
this idea to other clusters, we suggest that the magnitude of the
bright end of the blue straggler distribution may be an indicator of
when the phase of primordial binary burning occurred in clusters, and
may thus correlate with the dynamical properties of the cluster, and
the formation rate of other anomalous populations which require
stellar encounters \citep{B95}. If this scenario is correct, one might
expect that the present binary fraction of 47~Tuc should be
substantially lower than those of M80 and M3. The decrease in the
binary fraction could be a function of binary properties, such as
binary period or mass ratio, since the collisional cross section for
binary stars depends on both quantities.  The decrease in binary
fraction could also be function of radial distance from the core,
since we expect that binary stars at the center of the cluster will be
destroyed earlier than those further out.  Testing these ideas in
detail will require construction of complete models of the dynamical
history of the relevant clusters, including consideration of the
evolution of the binary population, an effort well beyond the scope of
this paper.

\section{SUMMARY}

We find that changes in the past formation rate of blue stragglers
produces drastic changes in their observed distribution in the CMD.  A
comparison between our parameterized models and observed blue
stragglers in 47~Tuc suggest that this cluster may have undergone an
epoch of enhanced BS formation several Gyrs ago.  We associate this
enhanced blue straggler formation rate with the epoch of primordial
binary burning invoked to explain the current characteristics of
several other clusters.  Since this epoch may well be short, it is
reassuring to find a cluster which has evidently gone through this
stage in the past, rather than experiencing it currently.  Much more
detailed dynamical models will be required to explore whether the
primordial burning scenario is consistent with the observed blue
straggler sequences in globular clusters.

\acknowledgements A. S. wishes to recognize support from the Natural
Sciences and Engineering Research Council of Canada. C.D.B. is supported
by NASA grant LTSA NAG5-6076.

\clearpage

\begin{figure}
\figurenum{1}
\plotone{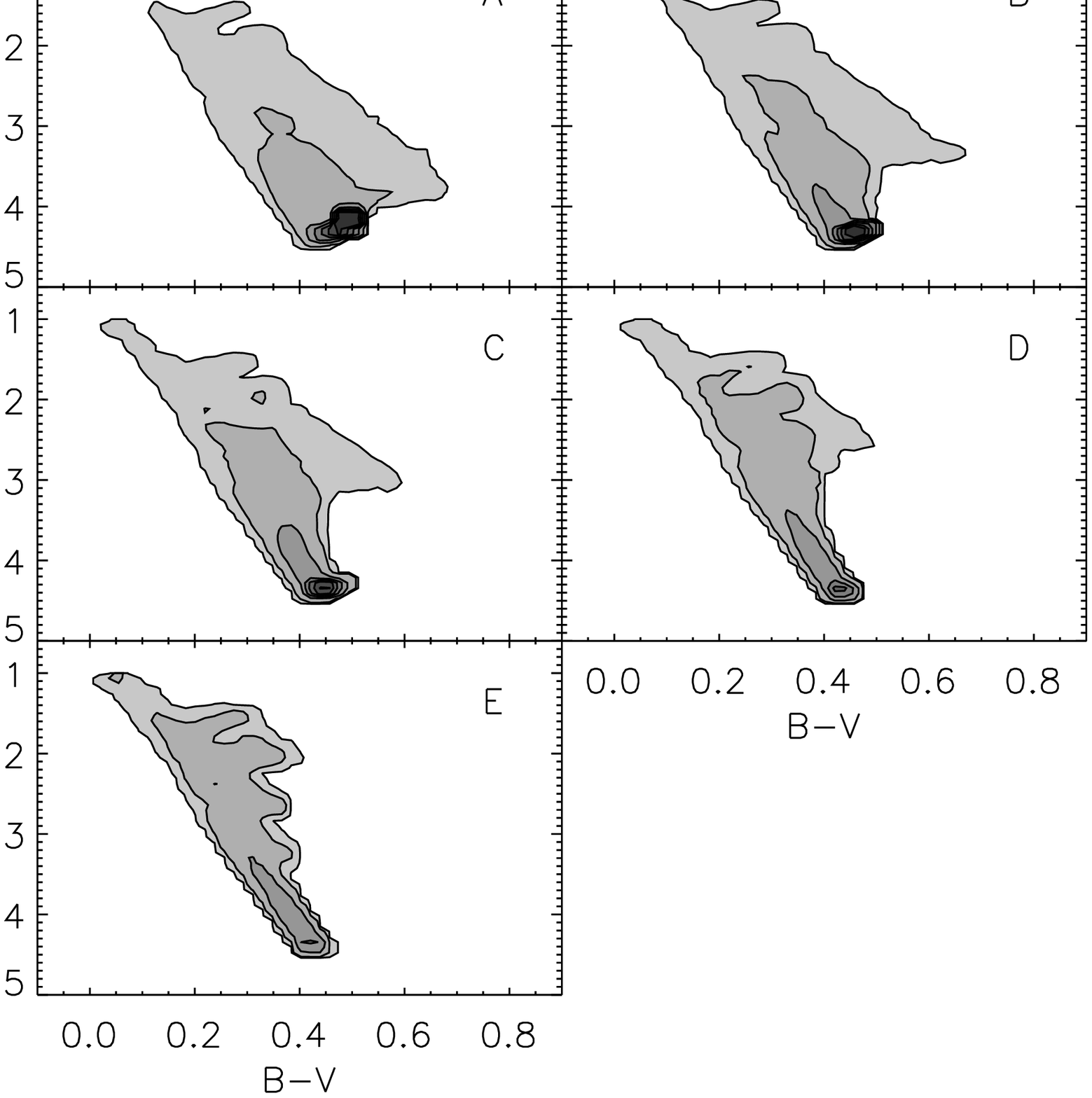}
\caption {Blue straggler distributions in the color magnitude diagram.
The contours represent density levels of stars per magnitude bin, with
darker colors corresponding to higher density. The blue straggler
formation rate is assumed to be constant, and lasting from some time
in the past to the current day. Panel A is the distribution which
results if blue stragglers are formed throughout the life of the
cluster. The blue straggler formation began 8 Gyr ago in panel B, 5
Gyr ago in panel C, 2 Gyr ago in panel D and 1 Gyr ago in panel E. The
contour levels are 0.1,1,5,10,15,20 and 25 blue stragglers per 0.012
magnitudes$^2$.}
\end{figure}

\begin{figure}
\figurenum{2}
\plotone{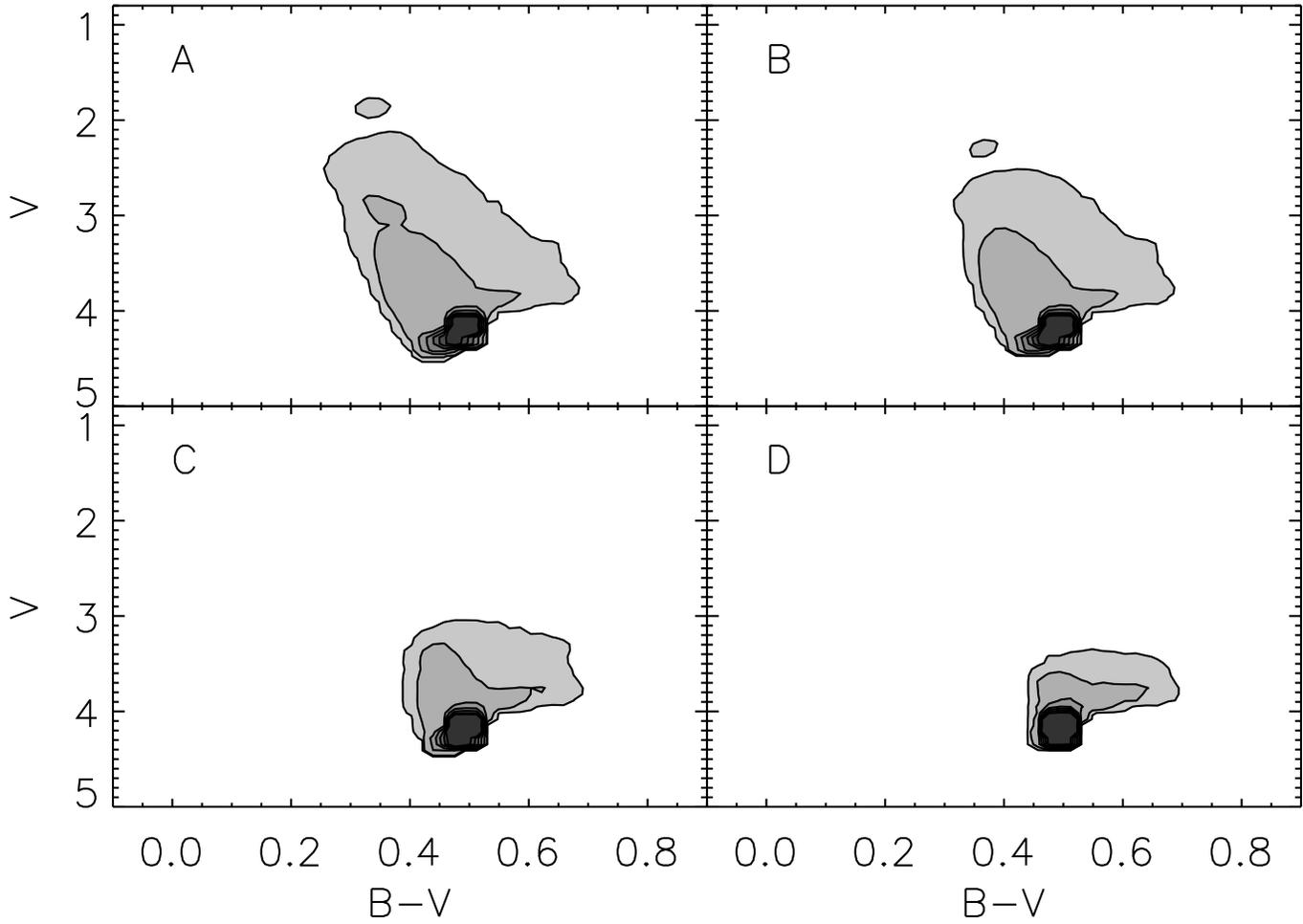}
\caption {Blue straggler distributions in the color magnitude
diagram. The blue straggler formation rate is assumed to be constant,
starting at the birth of the cluster, and ending some time in the
past: 1 Gyr ago (panel A), 2 Gyr ago (panel B), 5 Gyr ago (panel C) or
8 Gyr ago (panel D).}
\end{figure}

\clearpage

\begin{figure}
\figurenum{3}
\plotone{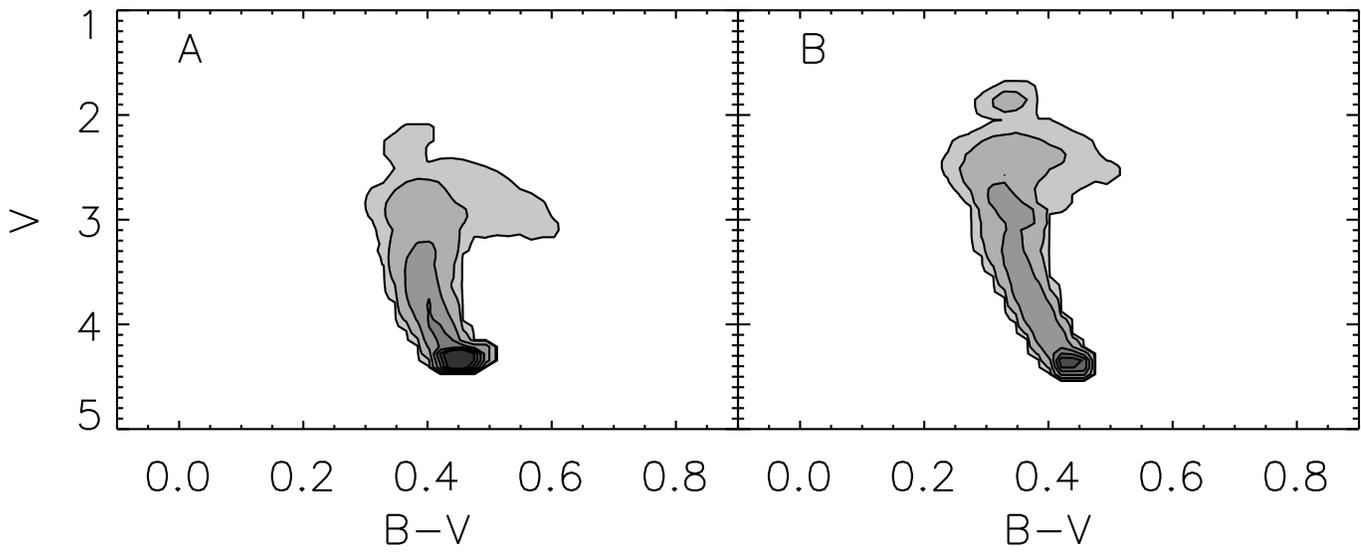}
\caption {Blue straggler distributions in the color magnitude
diagram. The blue straggler formation rate is assumed to be constant
between 2 and 5 Gyr ago (panel A), or between 1 and 2 Gyr ago (panel B).}
\end{figure}

\clearpage
\begin{figure}
\figurenum{4}
\plotone{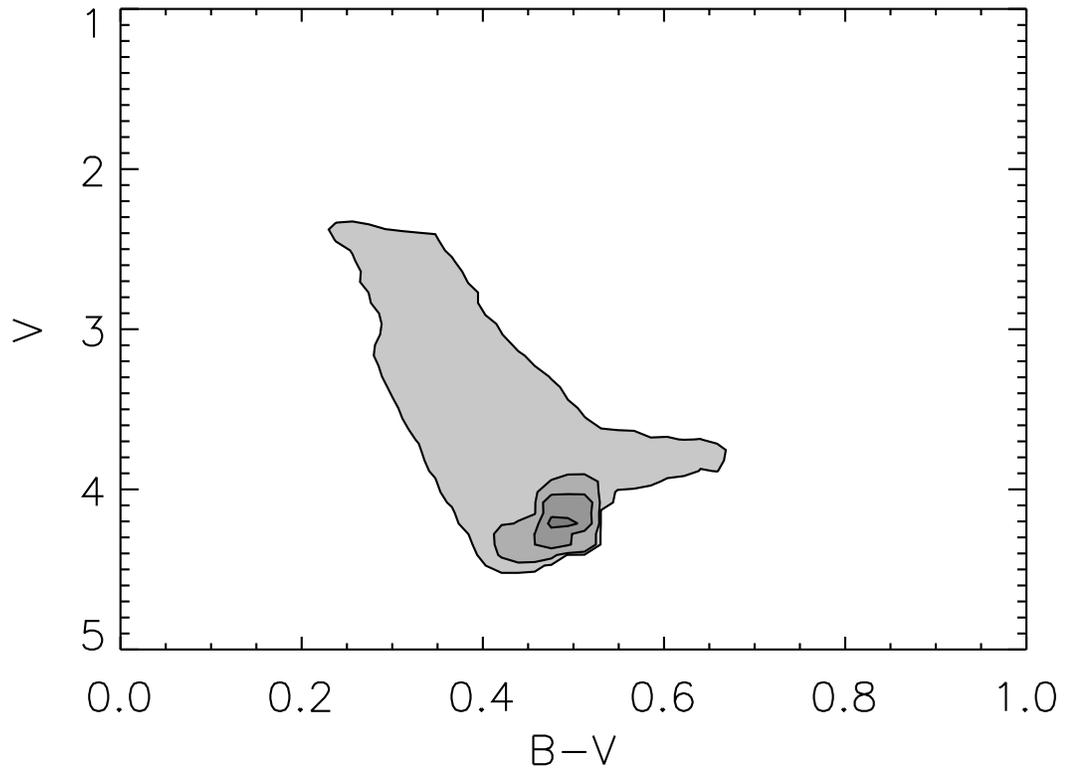}
\caption {Predicted distributions of blue stragglers in the
color-magnitude diagram of 47 Tuc. The blue straggler formation rate
is equal to the binary star destruction rate from \cite{hmr92}.}
\end{figure}

\clearpage

\begin{figure}
\figurenum{5}
\plotone{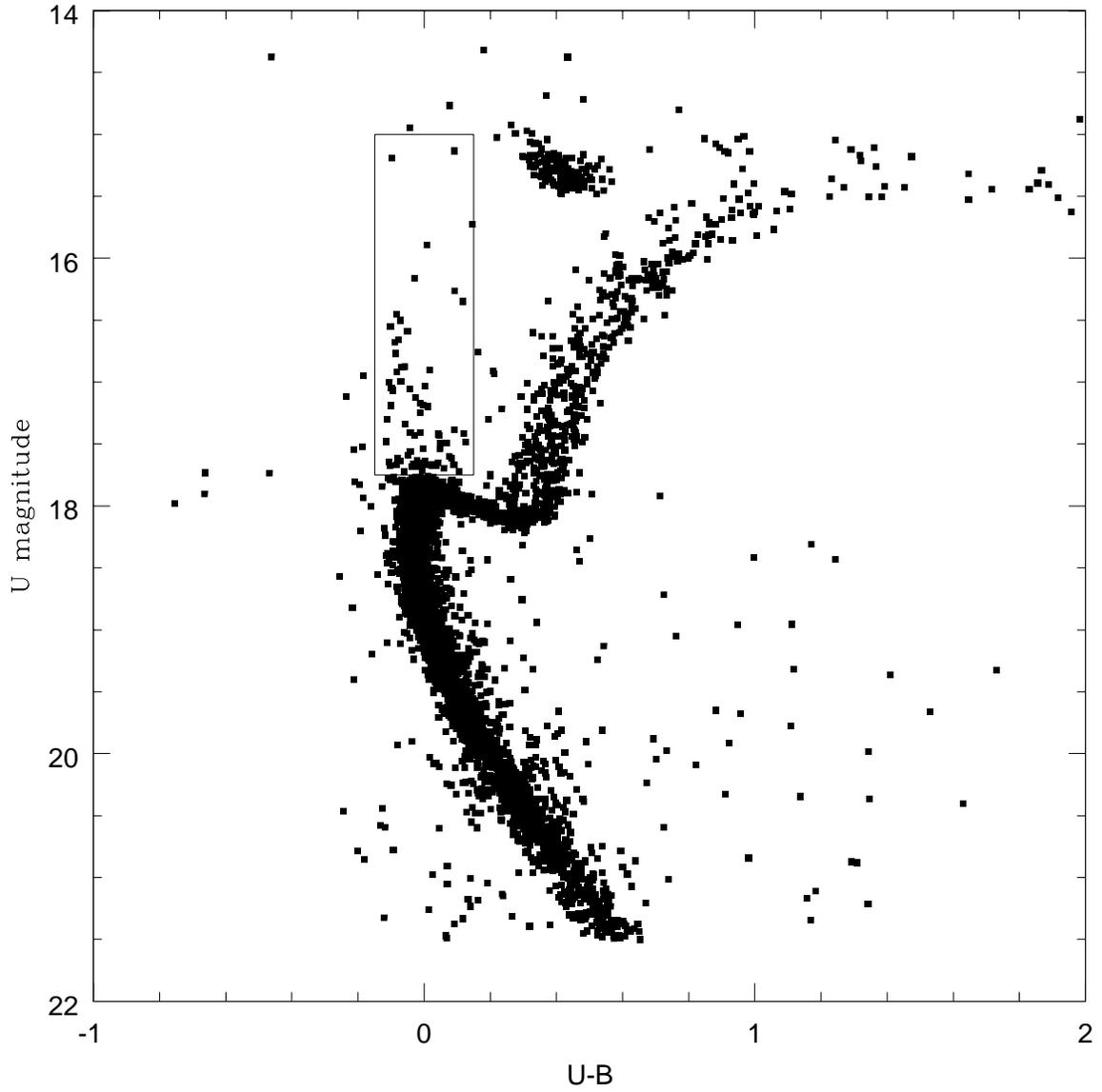}
\caption {The U vs. U-B color-magnitude diagram for 47 Tucanae,
showing all 4695 stars which contribute more than 50\% of the light
within 1 PSF radius. The box shows the first selection criterion for
blue stragglers.}
\end{figure}

\clearpage

\begin{figure}
\figurenum{6}
\plotone{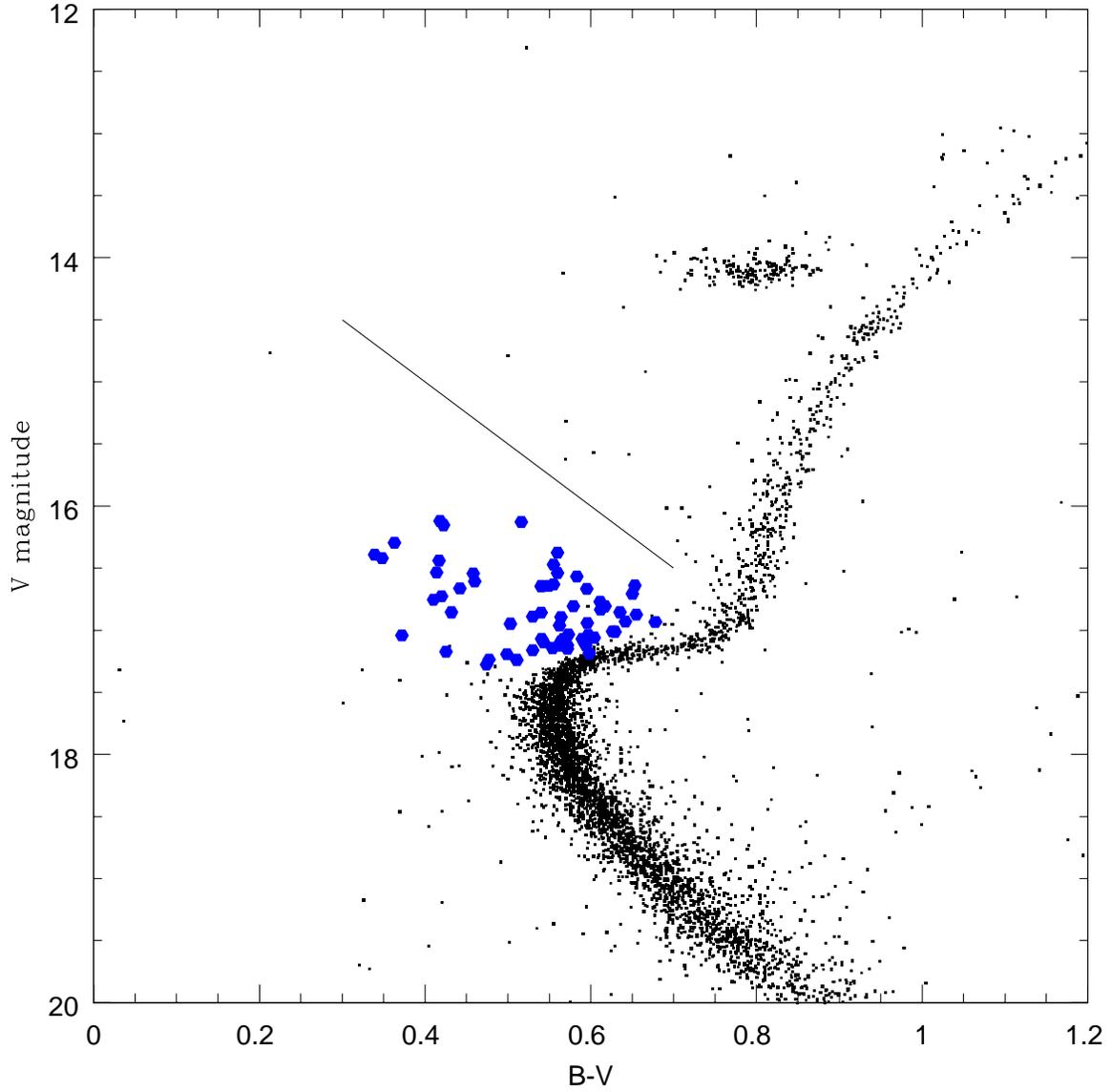}
\caption {The V vs. B-V color-magnitude diagram for 47 Tucanae. Blue
stragglers are constrained to lie below the solid line. The selected
sample of blue stragglers is shown as large points. }
\end{figure}

\clearpage

\begin{figure}
\figurenum{7}
\plotone{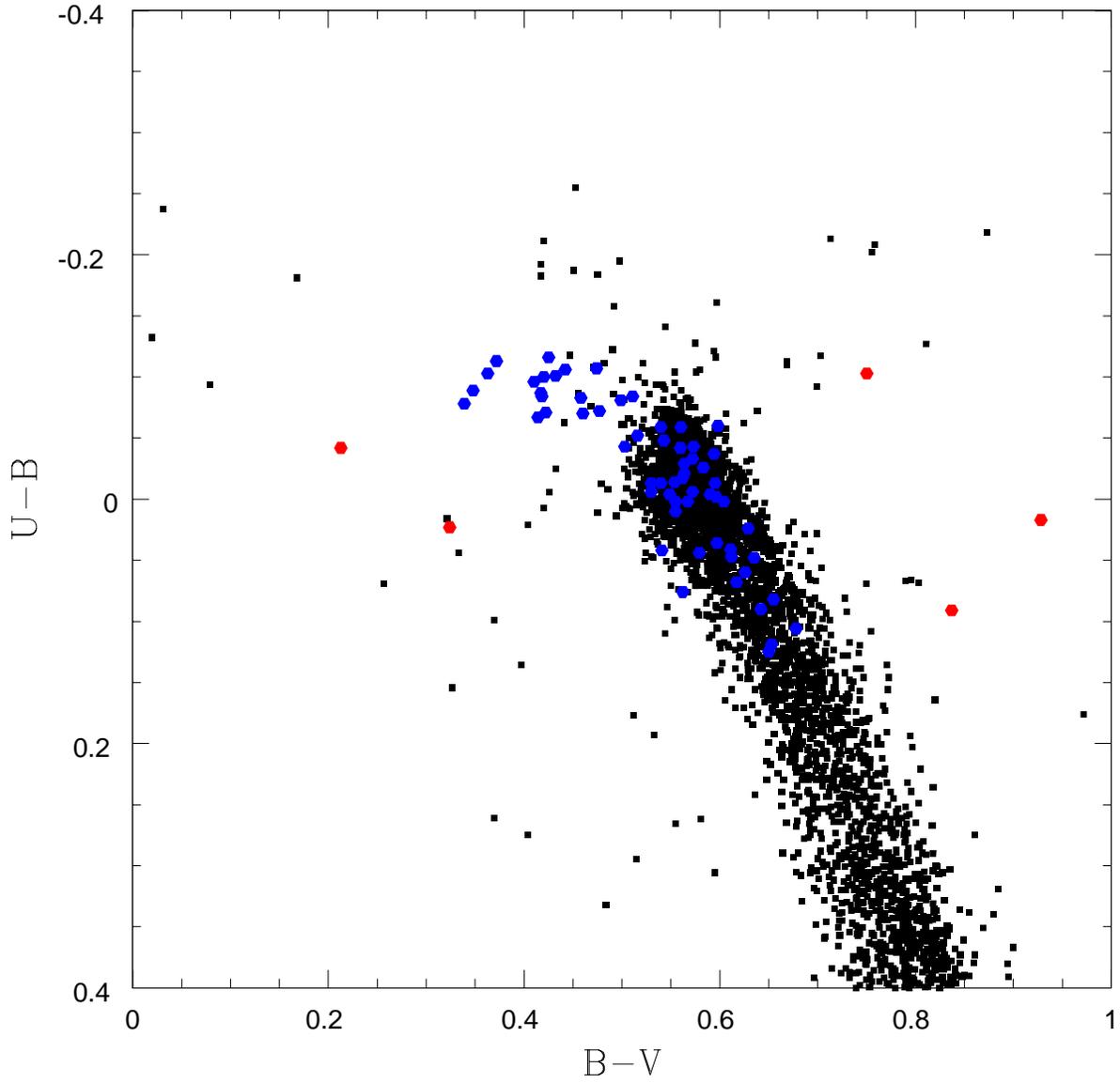}
\caption {The U-B vs. B-V color-color diagram for 47 Tucanae. The
selected sample of blue stragglers is shown as the blue symbols. The
5 red symbols which lie off the principal sequence of the cluster
were rejected from the sample. }
\end{figure}

\clearpage

\begin{figure}
\figurenum{8}
\plotone {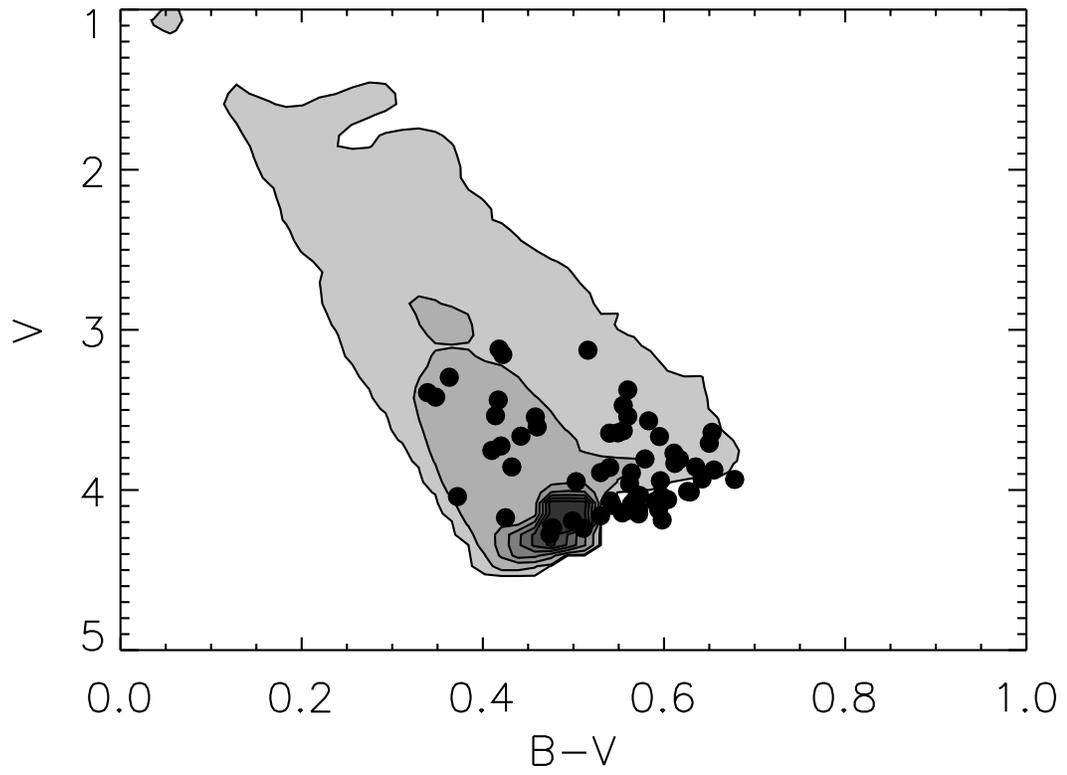}
\caption {Predicted distribution of blue stragglers in the
color-magnitude diagram of 47 Tuc. The blue stragglers were
formed at a constant rate throughout the life of the cluster. The
solid dots show the observed blue stragglers. }
\end{figure}

\clearpage

\begin{figure}
\figurenum{9}
\plotone{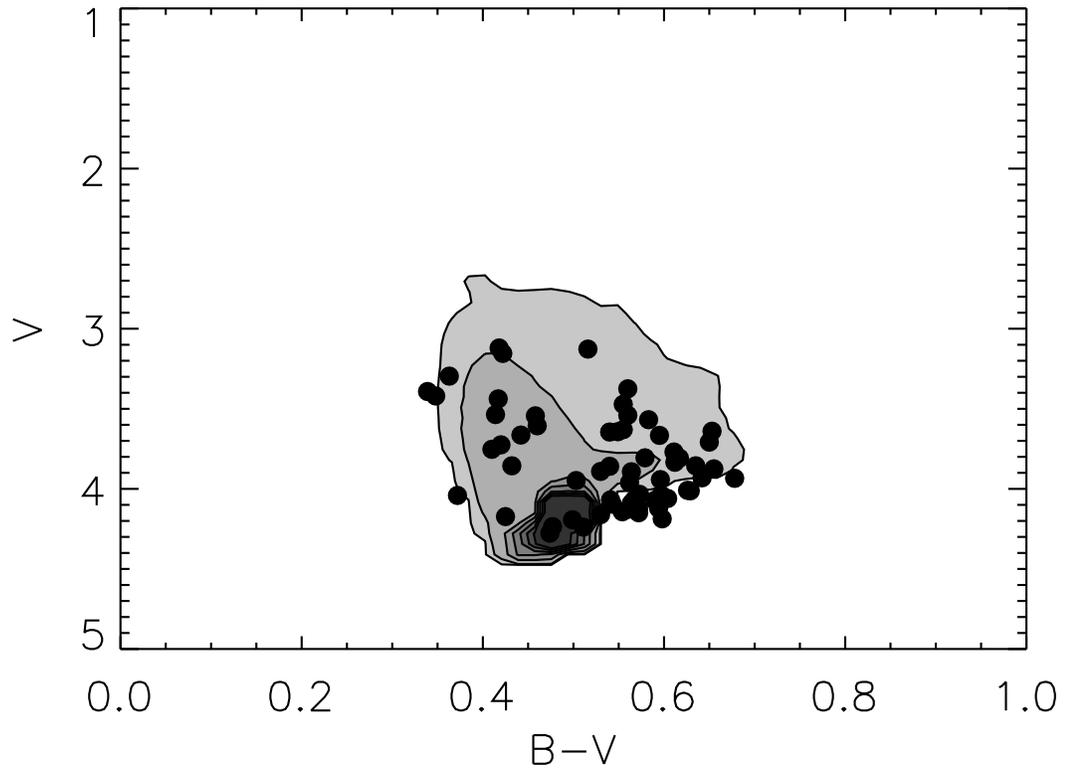}
\caption {Predicted distributions of blue stragglers in the
color-magnitude diagram of 47 Tuc. The blue stragglers were
formed at a constant rate until 3 Gyr ago, and no blue stragglers were
formed after that.}
\end{figure}

\end{document}